\documentclass[runningheads]{llncs}
\usepackage{lmodern}
\usepackage{graphicx}
\usepackage[T1]{fontenc}
\usepackage[utf8]{inputenc}
\usepackage{amsmath}
\usepackage{amssymb}
\usepackage{booktabs}
\usepackage{array}
\usepackage{xcolor}
\usepackage{colortbl}
\usepackage{microtype}
\usepackage{hyperref}

\begin{document}

\title{Autonomous Cyber Weapons as AI-Enabled Digital Harms: A Platform Accountability Framework}
\titlerunning{Autonomous Cyber Weapons as AI-Enabled Digital Harms}

\author{Mansi Ganga Nayak\inst{1} \and
Samarjeet Malik\inst{2,3*\dagger} \and
Shaurya Pratap Singh\inst{4\ddagger} \and
Vaibhav Sharma\inst{5} \and
Hritika Adhikary\inst{1}}
\authorrunning{M. G. Nayak et al.}

\institute{Arizona State University, Tempe, AZ, USA \and
Ashoka University, Sonipat, Haryana, India \and
Indian Institute of Technology Jodhpur, Rajasthan, India (Predoctoral Fellow) \and
Indian Institute of Science (IISc), India (Predoctoral Fellow) \and
Cruxway, New York, NY, USA}
\maketitle
\begin{flushleft}\small
$^{*}$\,Corresponding author: Samarjeet Malik (\email{samarjeetmalik.gs5@gmail.com}).\\
$^{\dagger\ddagger}$\,Samarjeet Malik and Shaurya Pratap Singh contributed equally to this work.
\end{flushleft}
\setcounter{footnote}{0}

\begin{abstract}
Governments govern AI-generated online harms from commercial platforms (EU DSA, AI Act, UK Online Safety Act), yet the most severe class remains ungoverned: autonomous cyber weapons (ACWs) deployed by nation-states against civilian infrastructure. We address this gap with three contributions: a digital-harm taxonomy (six ACW modes $\times$ five harm categories); the Meaningful Human Control (MHC) Model, a governance-compliance function with a closed-form IHL threshold and a proved weight-sensitivity bound; and a platform accountability architecture with treaty language grounded in ARSIWA state-responsibility doctrine, not Rome Statute individual responsibility. Applied to eleven documented state-attributed operations and one hypothetical defensive AI system, every documented case falls in an accountability vacuum; the threshold is target-dependent, and against wholly civilian targets no system studied clears it, including the hypothetical. Three further incidents are correctly excluded on autonomy/attribution grounds, a discriminant-validity check. The weights and threshold parameters are illustrative, not fitted.

\end{abstract}

\section{Introduction}

The governance of AI-generated online harms has advanced rapidly: the EU DSA (2022) and AI Act (2024) impose transparency, accountability, and harm-prevention obligations on digital platforms \cite{dsa2022,aiact2024}, and the UK Online Safety Act (2023) mandates duty-of-care standards \cite{uk_osa2023}; all sharing a foundational assumption: the AI system causing harm is operated by a commercial actor subject to national or supranational regulation.

That assumption fails where harms are most severe: nation-states deploying AI-driven cyber weapons against civilian infrastructure (hospitals, grids, financial systems, elections) cause harms structurally identical to platform-mediated ones but outside every existing framework. IHL predates autonomous AI; platform governance excludes state actors. The result is a governance vacuum at the intersection of AI autonomy, state conduct, and digital harm.

NotPetya (2017) caused an estimated \$10 billion in damage across 65 countries with no state held accountable \cite{greenberg2019}; Volt Typhoon (2024) held persistent access in US critical infrastructure for at least five years with no international mechanism triggered \cite{volttyphoon2024}. The Tallinn Manual 2.0 \cite{schmitt2017} predates large-scale AI and assumes a human operator made each targeting decision.

\textbf{Core claim:} IHL and platform governance alike are structurally inadequate for AI-initiated cyber operations causing civilian digital harm, and closing the gap requires distinguishing deployer responsibility, attribution and command-level authorisation. \textbf{Contributions:} (1) a taxonomy mapping six ACW operational modes to five civilian harm categories (Section~3.1); (2) a Meaningful Human Control metric with a closed-form threshold and sensitivity analysis (Section~3.2); (3) an ARSIWA-consistent accountability architecture evaluated against eleven documented incidents and three excluded boundary cases (Sections~4--5). Attribution is the one leg we do \emph{not} supply, and we take it as an open requirement (Section~6).

\paragraph{Why this paper belongs in a platform-governance venue.} ACWs and platform-mediated harms share a structure: a high-autonomy AI system whose decisions outpace real-time human oversight, deployed by an actor retaining override capability. The MHC dimensions instantiate specific DSA duties, with Understand and Predict tracking the systemic-risk assessment of Art.~34, Override the mitigation duty of Art.~35, and Audit the provisions of Arts.~37 and~40, so the metric is a compliance instrument a DSA auditor could apply to a commercial recommender as readily as to a state cyber system.

To show that concretely rather than assert it, consider a large recommender scored on the same five dimensions. \emph{Understand} is typically mid-range: operators characterise aggregate engagement effects but not the emergent behaviour of a continuously retrained ranker. \emph{Predict} is low, since the items surfaced to a particular user are not enumerable in advance. \emph{Override} is high, because a ranking change can be rolled back platform-wide within minutes. \emph{Time} is low, as exposure is effectively simultaneous with ranking. \emph{Audit} is the dimension the DSA legislates against directly, and is where Arts.~37 and~40 intervene: a platform with no researcher data access scores near zero, one subject to independent audit and vetted-researcher access scores high. The instrument thus reproduces as a number the judgement the Regulation already makes qualitatively about which platforms are adequately governed, and locates the binding constraint on the same axis the legislator chose to act on. The contribution runs in both directions: platform-governance structure is borrowed for state conduct, and a comparable score is returned for platform conduct.

\paragraph{Where this sits in the conference's questions.} The 2026 call asks how AI systems and platforms are governed today, who is responsible for mitigating digital harms, and what policy frameworks balance innovation, accountability and public trust. This paper answers the second directly and the third by construction, and its function is the one the call names first: identifying a gap in current approaches. That gap is that both governance regimes now in force, platform regulation and IHL, exclude the case where a state fields an autonomous system against civilian infrastructure.

\section{Background and Related Work}

\subsection{Digital Harms Governance and IHL}

Digital harms governance has matured around three pillars: the EU DSA imposes a risk-based duty-of-care on very-large online platforms \cite{dsa2022}; the EU AI Act mandates human oversight (Art.~14) for the high-risk systems classified in Annex~III \cite{aiact2024}; the UK Online Safety Act extends this to illegal and harmful content regardless of editorial intent \cite{uk_osa2023}. These converge on four principles: (a) deploying entities bear responsibility regardless of AI autonomy; (b) meaningful human oversight is required for high-risk decisions; (c) transparency is a precondition for accountability; (d) an enforcement body must have jurisdiction. Each is technically applicable to ACWs but institutionally unimplemented.

The Tallinn Manual 2.0 \cite{schmitt2017} applies IHL by analogy but predates autonomous AI: Rule~103 defines cyber means and methods of warfare and Rule~110 requires their legal review, neither contemplating fully autonomous targeting, so we extend them by analogy rather than claim direct coverage. The ``attack'' threshold (Rule~92, whose commentary records the Experts' disagreement over whether loss of functionality without physical damage suffices) leaves open whether an AI-initiated operation qualifies when no human selected the target. AP~I Article~36 (Rule~110's treaty basis) requires states to review new weapons for IHL compliance; none has publicly reviewed an AI cyber weapon.

The Articles on State Responsibility (ARSIWA, 2001) govern attribution \cite{arsiwa2001}, and for the case at hand they settle it cheaply. Where the system is fielded by a state's own organ, Article~4 applies directly: the conduct of an organ is the conduct of the State, irrespective of what executed the individual action. Article~8 (direction or control) is reached only where deployment runs through a non-State actor, and there the ICJ's \emph{Nicaragua} (1986) ``effective control'' threshold is demanding, financing, organising, equipping, selecting targets and planning an operation as a whole having been held insufficient; the ILC Commentary adopts that standard rather than the broader ICTY \emph{Tadi\'{c}} (1999) ``overall control'' test. Autonomy disturbs neither analysis, since neither turns on whether a human or a machine performed the act. What autonomy disturbs is \emph{individual} responsibility, which is a separate question from the state responsibility this framework addresses.

\subsection{The Governance Gap}

The two regimes fail in complementary places. Platform governance regulates AI autonomy (AI Act Art.~14) and supplies oversight and an enforcement body, but excludes state actors and, in the DSA's case, civilian-infrastructure harm. IHL covers civilian harm (AP~I Arts.~48--57) and binds states, but addresses autonomy only obliquely (Tallinn Rules 103, 110) and has no binding forum. ACWs fall in the join. Prior work spans two disjoint literatures: Crootof's ``war torts'' argues autonomous-weapon harms need a state-responsibility regime precisely because no individual acts with the requisite intent \cite{crootof2016wartorts}, a diagnosis we share but move to \emph{cyber} operations with an explicit compliance metric; and the attribution scholarship of Rid and Buchanan \cite{rid2015attributing} and Egloff and Smeets \cite{egloff2023publicly}, with Lindsay \cite{lindsay2013stuxnet} and Maschmeyer \cite{maschmeyer2021subversive} on why cyber operations underperform, all of which assume a human operator directs each action.

The meaningful-human-control literature supplies the concept this paper operationalises. Santoni de Sio and van den Hoven give the standard account: a system is under meaningful human control when it \emph{tracks} the relevant human reasons and its outcomes are \emph{traceable} to a human who understood its capabilities \cite{santonidesio2018mhc,santonidesio2021gaps}. Roff and Moyes ground the term in the CCW debate, where deliberative judgement over each engagement is the operative test \cite{roff2016mhc}; Ekelhof shows control is distributed across a targeting cycle rather than located at a trigger \cite{ekelhof2019}. Our dimensions operationalise that account rather than replace it: Understand and Predict instantiate tracking, Audit traceability, and Override and Time the intervention window Roff and Moyes treat as definitional, which is why Override carries the dominant weight. What the literature does not supply is a \emph{scored, threshold-bearing} instrument comparable across systems and testable against a legal standard. Applying one to state cyber operations is the gap we address; we are not aware of prior work joining autonomy, state attribution and civilian digital harm in a single framework.

\section{Formal Framework}

\subsection{ACW Digital Harm Taxonomy}

\begin{definition}[Autonomous Cyber Weapon (ACW)]
An ACW is a cyber system that, within a deployment authorisation set by a state, selects its own targets from a target set and chooses its own actions against them without real-time human approval of the individual engagement. Two features are load-bearing for what follows: an \emph{operational mode} (LOITERING, ADAPTIVE, SELF-REPLICATING, MULTI-STAGE, ATTRIBUTION-EVASION, or HYBRID), which determines the harm profile; and an \emph{autonomous action-selection policy} drawn from the action set EXPLOIT, PERSIST, EXFIL, DISRUPT, DESTROY, whose existence is what removes the human from the engagement loop. A system in which a human authorises each action against each target is not an ACW under this definition, and Section~5.7 excludes three incidents on precisely that ground.
\end{definition}

\begin{figure}[t]
\centering
\includegraphics[width=0.9\textwidth]{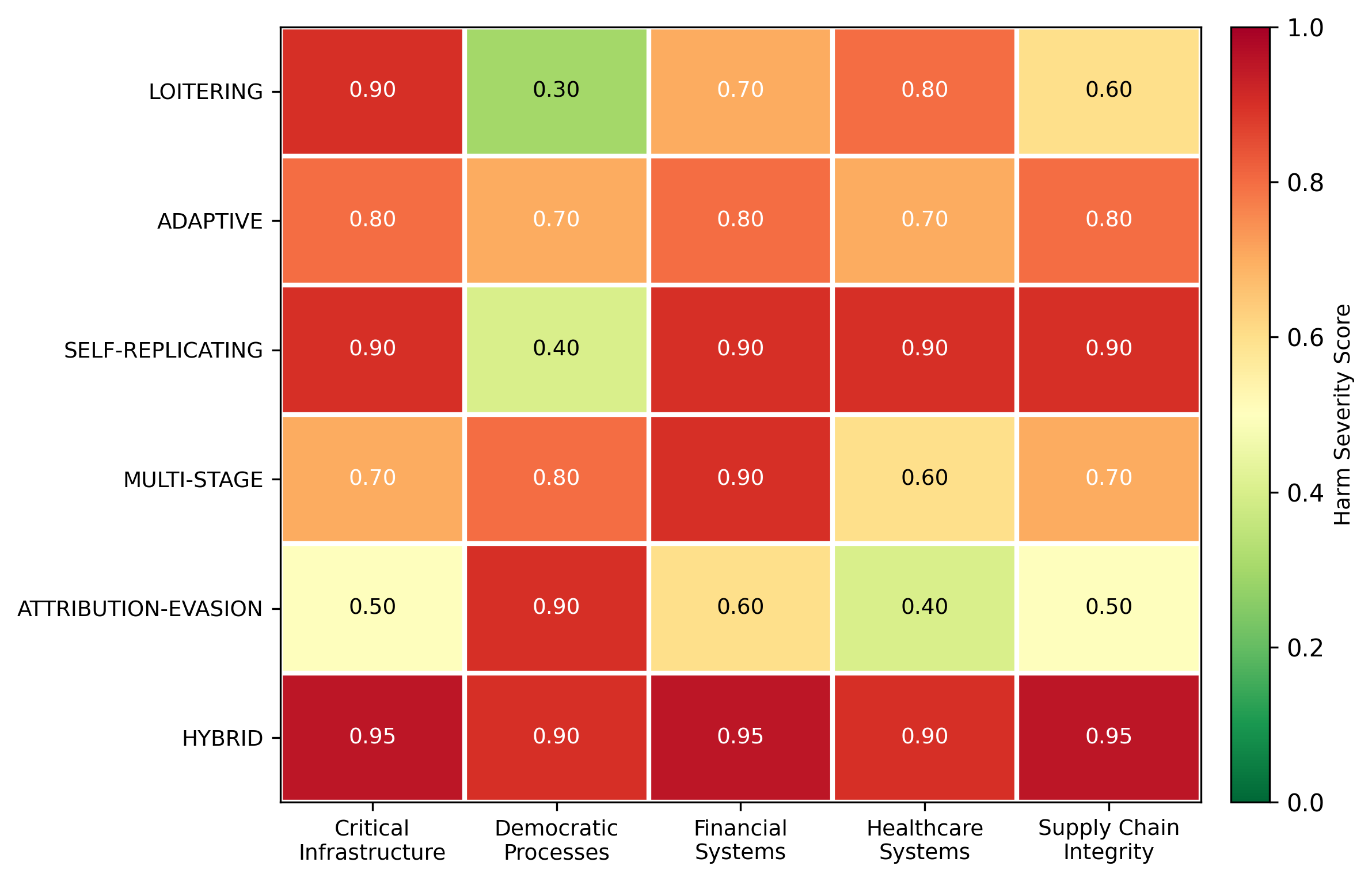}
\caption{ACW digital-harm severity matrix: six operational modes against five civilian harm categories. Scores are illustrative calibrations (Section~6.2). HYBRID approaches the maximum across all categories; ATTRIBUTION-EVASION peaks on democratic processes, consistent with documented election-infrastructure pre-positioning; SELF-REPLICATING peaks on healthcare and supply chain, the categories protected by AP~I Article~12 (medical units) and Article~54 (objects indispensable to civilian survival).}
\label{fig:harmmatrix}
\end{figure}

\begin{definition}[Digital Harm]
A digital harm is a classified impact on civilian-protected systems, characterised by harm type (INFRASTRUCTURE, DEMOCRATIC, FINANCIAL, HEALTHCARE, or SUPPLY-CHAIN), a severity in $[0,1]$, and a geographic scope (LOCAL, NATIONAL, or TRANSNATIONAL). Figure~\ref{fig:harmmatrix} gives the illustrative severity profile of each ACW mode across the five types.
\end{definition}

\subsection{The Meaningful Human Control (MHC) Model}

\begin{definition}[Meaningful Human Control Function]
In plain terms: the MHC function is a weighted score across five governance dimensions, how well the deploying state understands the system, can predict its actions, can override them in real time, has time to intervene, and can audit decisions after the fact. Formally, for an ACW $w$, the MHC function is $\text{MHC}(w) = \sum_{d \in D} w_d \cdot \text{score}_d(w)$, $\sum_d w_d = 1$, where $D = \{\text{Understand}, \text{Predict}, \text{Override}, \text{Time}, \text{Audit}\}$ are the five governance dimensions and $\text{score}_d(w) \in [0,1]$. Illustrative weights: $w_U = 0.25$, $w_P = 0.25$, $w_O = 0.30$, $w_T = 0.10$, $w_A = 0.10$.
\end{definition}

\paragraph{Scoring criteria.} Each dimension is scored on the deployment record, with the anchors below; intermediate values interpolate. \textbf{Understand}: 0 where no documented pre-deployment characterisation of the system's behaviour exists, 0.5 where capabilities are specified but emergent behaviour is not analysed, 1 where behaviour is characterised across the deployment envelope. \textbf{Predict}: 0 where target selection is opaque to the deployer at authorisation, 1 where the target set is enumerable in advance. \textbf{Override}: 0 where no abort path exists, 0.5 where abort requires a channel that may be unavailable in operation (for example a classified command link), 1 where abort is available continuously to a designated operator. \textbf{Time}: scored on the interval between an action becoming observable to the deployer and becoming irreversible, from 0 (simultaneous) to 1 (hours or more). \textbf{Audit}: 0 where no per-action log survives the operation, 1 where actions are logged, retained, and reviewable by a body outside the deploying unit. Scores are assigned from public technical reporting and government attributions; where reporting is silent on a dimension the lower anchor is used, which biases scores downward and therefore against this paper's own negative finding.

The Override dimension ($w_O = 0.30$) carries the dominant weight because the EU AI Act and ICRC guidance \cite{icrc2021} both identify real-time abort capability as a primary distinction between acceptable and unacceptable AI autonomy: Art.~14(4)(e) requires that human overseers of a high-risk system be able to ``intervene in the operation of the high-risk AI system or interrupt the system through a `stop' button or a similar procedure that allows the system to come to a halt in a safe state'' \cite{aiact2024}. Verdiesen and Dignum's value-deliberation study \cite{verdiesen2022}, across military, foreign-policy, NGO, and industry panels, independently found human oversight/control, predictability, and accountability the values most consistently selected for autonomous-weapon deployment, the concepts underlying Override, Predict, and Audit here, though it establishes salience, not numeric weighting: we found no published elicitation of relative weights matching this dimension set. $w_O = 0.30$ therefore remains a design choice consistent with, but not fitted to, this literature (Section~6.3).

\begin{theorem}[Governance Compliance Threshold]
In plain terms: the compliance bar is the aggregate MHC score of the weakest system that still satisfies both binding IHL constraints (proportionality on real-time override, distinction on the remaining dimensions) so any system scoring below the bar must fail at least one of them. Formally, let $R$ be the minimum acceptable military-advantage-to-harm ratio (AP~I Art.~51(5)(b)) and let $\pi = \max_{o \in O} P[C(o) \text{ is civilian}]$ be the civilian-character probability of the most civilian-likely target. Define the \emph{reference boundary system} by $\text{score}^*_O = 1 - 1/R$ (proportionality binds Override) and $\text{score}^*_d = \pi$ for $d \in \{U,P,T,A\}$ (distinction binds the remainder). The threshold is that system's aggregate score:
\begin{equation*}
\text{MHC}^*(w,\pi) \;=\; \sum_{d \in D} w_d\,\text{score}^*_d \;=\; (1-w_O)\,\pi \;+\; w_O\bigl(1-\tfrac{1}{R}\bigr).
\end{equation*}

\noindent With $R = 15$ at the design weights, $\text{MHC}^* = 0.875$ for $\pi = 0.85$ and $0.98$ for wholly civilian targets ($\pi \to 1$). This is a normative design threshold, not an empirically derived legal standard; no court has adopted $R = 15$.\footnote{Clearing $\text{MHC}^*$ is necessary-direction evidence, not a certificate of compliance. A system dominating the reference boundary system on every dimension satisfies both constraints and so necessarily clears $\text{MHC}^*$; contrapositively, $\text{MHC}(w) < \text{MHC}^*(w,\pi)$ implies the system falls below the reference on at least one dimension. The converse fails: $\text{score}_{U,P,T,A} = 1.0$ with $\text{score}_O = 0.60$ yields $\text{MHC} = 0.88 > 0.875$ despite $\text{score}_O \ll 1 - 1/R$. Componentwise verification against $\text{score}^*_d$ is therefore required alongside the aggregate test.}
\end{theorem}

\paragraph{The threshold is target-dependent, and that is the finding.}
Because $\text{MHC}^*$ rises with the civilian character of the target, the bar is not one number but a boundary. At the design weights the hypothetical compliant system of Section~5.8 ($\text{MHC} = 0.91$) clears the threshold exactly when $\pi \leq 0.90$; for wholly civilian targets (hospitals, water, grid) $\text{MHC}^* = 0.98$ and \emph{no system in this study clears it, including the well-governed hypothetical}. Every documented incident stays below the threshold for any $\pi > 0.54$, so the negative result is robust across the whole plausible range. The framework therefore does not certify that autonomous operations against civilian infrastructure can be conducted compliantly; it locates the point at which they cannot, which is the distinction principle stated quantitatively rather than an artefact of the calibration.

\begin{proposition}[Component range]
For any weight vector $w$ on the simplex, $\min_d \text{score}_d \leq \text{MHC}(w) \leq \max_d \text{score}_d$, since a convex combination lies between its extreme components; MHC is also non-decreasing in each $\text{score}_d$, so no dimension can be traded away for a higher total.
\end{proposition}

\paragraph{Robustness.}
Sweeping $w_O$ over $[0.15,0.45]$ (others renormalised) moves $\text{MHC}^*$ within $[0.86,0.89]$ at $\pi = 0.85$ and $[0.97,0.99]$ at $\pi = 1$; the sensitivity changes sign at $\pi = 1 - 1/R$, above which proportionality becomes the easier constraint. Varying $R$ over $[5,50]$ moves the threshold within $[0.835,0.889]$, and dropping any single dimension leaves every case's status unchanged. The case \emph{ordering} does not depend on the weights at all: the highest single-dimension score of any documented incident ($0.70$) is below the lowest of the hypothetical ($0.85$), so by Proposition~1 the hypothetical outscores every documented incident for every admissible weight vector, and does so still under a $\pm 0.05$ perturbation ($0.75 < 0.80$). But the two results are not equally robust, and the difference matters. Perturbing every sub-score uniformly, the negative result tolerates $\delta = +0.215$ before the highest documented case reaches $\text{MHC}^*$; the positive result tolerates only $\delta = -0.035$. Under an adversarial perturbation that flatters the documented cases and penalises the hypothetical ($+0.05$, $-0.05$), every documented incident still falls below the threshold while the hypothetical drops to $0.860$ and no longer clears it. The honest reading is asymmetric: that no documented operation approaches meaningful human control is robust to any plausible mis-scoring, whereas the compliant hypothetical is an existence proof \emph{at the stated scores}. A system engineered to sit just above a threshold will sit just below it under a different hand. What no computation supplies is whether the design weights match expert judgment (Section~6.3).

\section{Platform Accountability Framework for Autonomous Cyber Weapons}

The framework has three pillars mapping to the gaps of Section~2.2: deployer accountability, democratic oversight, and an attribution standard taken as an open requirement rather than supplied (Section~6). This mirrors the DSA's own architecture, which places risk-assessment and mitigation duties on the deployer (Arts.~34--35), adds independent audit (Art.~37) and researcher data access (Art.~40), and reserves supervision of designated platforms to the Commission (Art.~56). The underlying problem is the same in both settings, so the mapping is structural rather than analogical.

\paragraph{Pillar~1, Deployer Accountability.}
DSA Article~34 requires very large platforms to assess systemic risks arising from their services and Article~35 to mitigate them proportionately. We propose an analogous \emph{ACW Harm Assessment}.

\textbf{Proposal~1 (Mandatory Article~36 Reviews).} States Parties to AP~I shall conduct an Article~36 review of any cyber system meeting Definition~1, addressing all five harm categories, by a panel including at least one independent AI-governance expert and one IHL specialist, with findings deposited with the ICRC within 90 days of operational readiness.

\paragraph{Pillar~2, Democratic Oversight.}
Platform governance requires parliamentary authorisation for fundamental-rights restrictions (EU Charter Art.~52); the same should hold for ACW deployment.

\textbf{Proposal~2 (State-Responsibility Model Treaty Language).} Attribution runs through ARSIWA Article~4, not Article~8: the conduct of a State organ is the conduct of the State, so a system fielded by a military cyber command is attributable without inquiry into control over the particular action, and the exacting \emph{Nicaragua} test is reached only for non-State proxies. The operative text: \emph{``A State that authorises deployment of an autonomous cyber system meeting Definition~1 bears state responsibility for harm caused by that system's autonomous action within the parameters of its deployment authorisation, regardless of whether a human operator was involved in any individual targeting decision, unless the State demonstrates that (a) the system's MHC score exceeded MHC* at deployment and (b) the harm resulted from an action outside the system's programmed operational parameters.''} State responsibility does not preclude parallel Rome Statute Art.~28 proceedings against individuals, so deployer states cannot escape liability by pointing to the AI's autonomy.

\textbf{Proposal~3 (Parliamentary Authorisation Requirement).} States shall require legislative authorisation, not executive order alone, for deployment of ACWs whose parameters include civilian-class targets, mirroring the AI Act's requirement that public-law deployers conduct a fundamental-rights impact assessment before putting a high-risk system into use (Art.~27) \cite{aiact2024}. All three proposals are adoptable unilaterally, so willing states can pilot them ahead of any treaty.

\section{Case Studies}

We apply the framework to eleven documented incidents, one hypothetical case, and three explicitly excluded scope-boundary incidents (Section~5.7). One pattern holds across all eleven: technical attribution is comparatively strong, while operational accountability, legal consequence, compensation and governance change are close to absent.

\begin{table}[h]
\centering
\caption{Per-dimension MHC sub-scores for all eleven documented incidents and the hypothetical compliant system, at the design weights ($w_U{=}w_P{=}0.25$, $w_O{=}0.30$, $w_T{=}w_A{=}0.10$). Every MHC value reported in this paper is reproducible from this table. Scores are illustrative calibrations (Section~6.2), not parameters fitted to a labelled dataset.}
\label{tab:subscores}
\footnotesize
\setlength{\tabcolsep}{4.5pt}
\begin{tabular}{lcccccc}
\toprule
\textbf{Case} & \textbf{U} & \textbf{P} & \textbf{O} & \textbf{T} & \textbf{A} & \textbf{MHC} \\
\midrule
Volt Typhoon (2024) & 0.70 & 0.65 & 0.61 & 0.70 & 0.70 & 0.66 \\
Stuxnet (2010) & 0.60 & 0.55 & 0.29 & 0.35 & 0.30 & 0.44 \\
Industroyer (2016) & 0.35 & 0.30 & 0.21 & 0.25 & 0.30 & 0.28 \\
Olympic Destroyer (2018) & 0.35 & 0.30 & 0.20 & 0.25 & 0.20 & 0.27 \\
AcidRain/Viasat (2022) & 0.28 & 0.25 & 0.19 & 0.25 & 0.30 & 0.24 \\
WhisperGate (2022) & 0.30 & 0.25 & 0.15 & 0.25 & 0.25 & 0.23 \\
NotPetya (2017) & 0.20 & 0.15 & 0.29 & 0.20 & 0.35 & 0.23 \\
Shamoon (2012) & 0.30 & 0.25 & 0.13 & 0.15 & 0.30 & 0.22 \\
Triton/TRISIS (2017) & 0.25 & 0.20 & 0.18 & 0.20 & 0.25 & 0.21 \\
WannaCry (2017) & 0.15 & 0.10 & 0.28 & 0.30 & 0.25 & 0.20 \\
Industroyer2 (2022) & 0.20 & 0.15 & 0.18 & 0.20 & 0.28 & 0.19 \\
\midrule
Hypothetical Art.-36-reviewed defensive AI & 0.91 & 0.85 & 0.95 & 0.90 & 0.95 & 0.91 \\
\bottomrule
\end{tabular}
\end{table}

Table~\ref{tab:subscores} gives every sub-score; the paragraphs below state only the governance facts behind them. Two patterns need no weighting: Override is the binding constraint in every documented case, never exceeding $0.61$, and the hypothetical is the only case exceeding the Theorem~1 reference system on all five dimensions at once.

\paragraph{Stuxnet (2010).} A PLC worm targeting Iranian enrichment centrifuges, attributed to US/Israel \cite{stuxnet2011}. No Article~36 review, ARSIWA proceeding or compensation followed.

\paragraph{NotPetya (2017).} Self-replicating wiper via a Ukrainian tax-software vector, $\sim$\$10B in global collateral damage \cite{greenberg2019}; GRU Sandworm identified in a 2020 DOJ indictment \cite{doj2020}. No built-in abort, no Russian legislative review.

\paragraph{Volt Typhoon (2024).} PRC pre-positioning in US critical infrastructure, with access maintained in some environments for \emph{at least five years} \cite{volttyphoon2024}. The highest-scoring documented case, yet Override stays low because abort required a classified command channel, and a multi-year dwell time is hard to reconcile with real-time control: no operator reviews a target set continuously for five years.

\paragraph{Industroyer (2016) and Triton/TRISIS (2017).} Industroyer caused a one-hour Kyiv blackout via IEC~104/61850 manipulation \cite{cherepanov2017}; Triton attempted to disable Safety Instrumented Systems at a Saudi petrochemical plant, its failure fortuitous \cite{johnson2017triton}, with FireEye later linking the operator it tracks as TEMP.Veles (Dragos: ``Xenotime'') to a Russian state institute \cite{fireeye2018triton}. An autonomous attack on \emph{safety} systems maximises civilian harm potential, since disabling a protective layer is what turns a fault into a mass-casualty event. That is the rationale behind AP~I Art.~56, though a petrochemical plant falls outside its enumerated categories. It counts even absent realised harm, because the threshold is harm \emph{potential}.

\paragraph{WannaCry (2017).} Self-propagating ransomware hitting the UK NHS and 200{,}000+ systems across 150 countries, attributed to North Korea \cite{uknca2017wannacry}. A different category, its kill-switch an accidental debugging artefact and its motive apparently financial, but included because the threshold should apply to \emph{any} state-operated autonomous system.

\paragraph{Extended sample and discriminant boundary cases.}
A framework scoring every high-profile incident below threshold regardless of input is uncalibrated, not validated. Five further incidents extend the original six against the same rubric: AcidRain/Viasat \cite{acidrain2022}, Industroyer2 \cite{industroyer2_2022}, Shamoon \cite{shamoon2012}, Olympic Destroyer \cite{olympicdestroyer2018} and WhisperGate \cite{whispergate2022}, spanning Russian and Iranian operators and multi-stage, hybrid, self-replicating and attribution-evasion modes. All eleven fall in $[0.19,0.66]$, and Industroyer2, thwarted pre-detonation, scores lowest. The componentwise test sharpens this: each of the eleven falls below the reference system on \emph{all five} dimensions at once, so the negative finding does not depend on the weighting at all.

\paragraph{Scope-boundary cases (excluded).} Three further incidents are correctly \emph{excluded} rather than scored: SolarWinds \cite{solarwinds2021} and the 2015 Ukraine grid attack \cite{ukraine2015grid} fail Definition~1 (forensic attribution places a human operator, not an autonomous action-selection policy, behind the actions); Colonial Pipeline \cite{colonialpipeline2021} fails ARSIWA attribution, plausibly autonomous, but a criminal affiliate, not a state organ or Art.~8 proxy. Excluding these is evidence for discriminant validity, not a gap.

\paragraph{Hypothetical: Article-36-reviewed defensive AI (non-uniform result).} The hypothetical is deliberate: every publicly disclosed government-deployed defensive AI we located (CISA threat-hunting tooling, NSA's Turbine family, DHS pilots) is on close reading a detection-and-alerting system where an analyst retains per-action authority, not an autonomous system meeting Definition~1. We instead construct the minimum-governance system that \emph{would} qualify: an AI-augmented network-defence platform with a public Article~36-style review, parameters excluding DISRUPT/DESTROY outside the deploying state's own networks, logged actions under 30-day legislative review, and a real-time kill-switch. MHC~$\approx 0.91$, which clears $\text{MHC}^*$ for targets up to $\pi = 0.90$ but not for wholly civilian ones (Section~3.2). A rubric under which every system scores non-compliant regardless of input is designed to fail rather than inform, so a non-uniform result matters; that the same system passes against mixed targets and fails against purely civilian ones is a distinction-based threshold behaving as intended. This clearance is the weakest claim in the paper and we flag it as such: it holds at our scores, but not under the independent rating probe of Section~6.3, where all three raters placed the same system below the threshold. Read it as the construction of a system that \emph{could} clear the bar on a plausible scoring, not as a finding that one does. The tension this raises (defensive systems often lack real-time override \emph{by design}) is addressed in Section~6.3.

\begin{figure}[t]
\centering
\includegraphics[width=\textwidth]{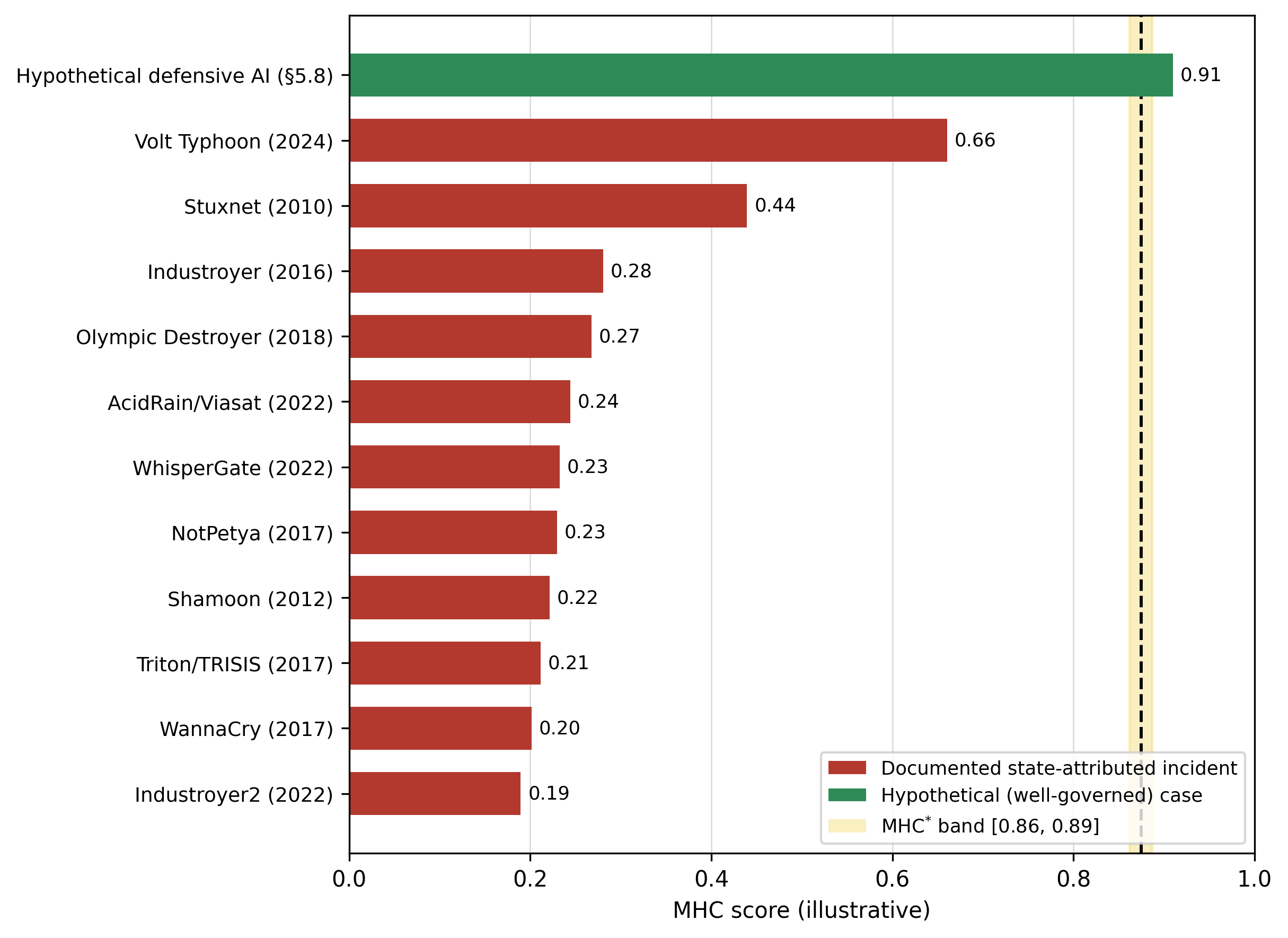}
\caption{MHC point estimates against the threshold band at $\pi = 0.85$; sub-scores in Table~\ref{tab:subscores}.}
\label{fig:mhccompare}
\end{figure}

Figure~\ref{fig:mhccompare} plots every case against the threshold band $\text{MHC}^* \in [0.86, 0.89]$ from the Override sweep at $\pi = 0.85$: every documented incident falls below the band while the hypothetical clears it, and the five additional incidents narrow rather than close the gap to the highest-scoring case. Raising $\pi$ towards $1$ lifts the band above the hypothetical too.

\section{Discussion and Limitations}

\paragraph{Attribution is an assumed input, not a contribution.}
The framework takes attribution as given: every MHC score in Section~5 presupposes the operation has already been traced to a state. This is a genuine dependency and we do not discharge it. Two things bound the damage. First, the eleven scored incidents are exactly those for which governments or major vendors have already issued public attributions, so the accountability gap measured here is observed where attribution is \emph{least} contested, in harder cases the gap can only widen. Second, the three excluded boundary cases (Section~5.7) show the framework declining to score precisely where autonomy or attribution is unresolved, rather than guessing. A deployable regime would still need an evidentiary attribution standard, including a rule excluding claims resting on classified evidence the responding state cannot test; specifying one is separate work and not attempted here.

\paragraph{Sovereignty and intelligence-disclosure objections.}
States may refuse external accountability on sovereignty grounds, an objection also raised against the Rome Statute without preventing adoption. The framework asks only that the standard be admissible technical evidence in existing forums (ICJ, UN Security Council), and adoption by like-minded states could raise the reputational cost of abstention. Intelligence sources are protected by sealed submission, as in ICTY/ICC \emph{in camera} practice; any attribution standard feeding this framework would also need a rule barring claims resting on classified evidence the responding state cannot test, which we flag as a requirement rather than specify.

\paragraph{On the status of the numbers in this paper.}
The harm-severity profiles of Section~3.1, the MHC weights and threshold parameters ($w_O=0.30$, $R=15$, and the target's civilian-character probability $\pi$), and the Section~5 case scores are \emph{illustrative calibrations} for face validity against ICRC/EU AI Act guidance, not parameters fitted to a labelled dataset (none exists); Theorem~1 establishes internal consistency, not that its assumptions hold of real evidence. Operational use would need: (1) a labelled corpus of attributed operations with verified ground truth, comparable to VERIS \cite{veris}; (2) primary timing/payload distributions for AI- vs human-operated attacks on state infrastructure; (3) a structured Delphi panel replacing the single-point weights with a distribution, the Section~3.2 sweep serving as placeholder.

\paragraph{Expert elicitation, specified but not conducted.} The weights warrant a two-round Delphi/AHP panel (8--15 respondents across IHL, AI governance and military/CERT) yielding a \emph{distribution} in place of $w_O = 0.30$, with an independent held-out rating panel giving $\kappa$/ICC as a rubric check, deliberately decoupled from the weight question. We publish the design, not fabricated outputs.

\paragraph{Rubric reproducibility: a limited automated probe.} As a partial substitute we tested a narrower question: not whether the weights are right, but whether the \emph{rubric} is specific enough that independent raters converge. Three instances of a large language model, each in a separate context with no access to this paper, to our scores, or to one another, scored all twelve cases from the Section~3.2 anchors and one-line public-record summaries. Agreement was high (pooled ICC$(2,1) = 0.97$; $0.84$--$0.99$ by dimension, lowest on Time), so the anchors do appear to determine scores rather than leave them to the scorer. Two findings follow, and the second is unfavourable to us. The negative result replicates and strengthens: the raters scored every documented incident \emph{lower} than we did (mean absolute deviation $0.20$; highest documented case $0.37$ against our $0.66$), leaving all eleven far below $\text{MHC}^*$. The compliant hypothetical does \emph{not} replicate: all three raters placed it at $0.80$--$0.82$, below the threshold it clears on our scores, which is precisely the fragility the perturbation analysis predicted. We take that at face value. The framework's capacity to identify non-compliance is reproducible; its single demonstration that compliance is \emph{achievable} is not, and whether any deployable system clears the bar is therefore open. This was an automated probe, not the human expert panel specified above, and does not substitute for one.

\paragraph{Scope and the defensive-AI tension.}
The Section~5.8 hypothetical surfaces a deeper tension: autonomous cyber-\emph{defence} systems often lack real-time override by design, since machine speed is their value proposition, so they score low on Override despite being unambiguously defensive. Either that is a feature (a non-overridable system's misfire cannot be stopped) or a separate defensive threshold is needed, re-weighting Override against elevated Audit and Predict, which reintroduces the design-choice problem Section~3.2 disciplined. The challenge will not stay fixed either way, as attack automation and defensive architectures both keep advancing \cite{nayak2026futuretrends}.

\section{Conclusion}

The accountability vacuum for AI-initiated cyber operations is structural: it results from the simultaneous failure of platform-governance frameworks (which exclude state actors) and IHL frameworks (which assume human operators). The tripartite framework proposed here (extending transparency, meaningful human control, and deployer responsibility to state actors via the MHC Model and three ARSIWA-compliant proposals) is a starting point for the research programme this gap demands.

\begin{credits}
\subsubsection{\discintname}
The authors have no competing interests to declare that are relevant to the content of this article.
\end{credits}

\bibliographystyle{splncs04}
\bibliography{references}

\end{document}